# Half-Integer Flux Quantization in Unconventional Superconductors

*C.C. Tsuei, IBM Thomas J. Watson Research Center, USA*
*J.R. Kirtley, Center for Probing the Nanoscale, Stanford University, USA*

**Integer flux quantization** is one of the most striking demonstrations of the macroscopic phase coherence of the charge-carrier pair wavefunction in superconductors. Based on the BCS theory, the ensemble of Cooper pairs can be described by a single complex wavefunction of electron wavevector **k** :

$$\Psi(\mathbf{k}) = |\Psi(\mathbf{k})| e^{i\varphi(\mathbf{k})} \qquad (1)$$

The amplitude $|\Psi(\mathbf{k})|$ determines properties like $T_c$, and the size of the BCS energy gap, $|\Delta(\mathbf{k})|$, which is a measure of pairing strength. The well-known hallmarks of superconductivity, such as zero resistance and the Meissner effect, are all derived from the phase $\varphi(\mathbf{k})$ and quantum phase coherence established among all the Cooper pairs on a global scale. The phenomenon of **integer flux quantization** is a consequence of the single-valueness of the pair wavefunction, or equivalently of the phase winding in multiples of $2\pi$ around any singly-connected path in a superconductor. The magnetic flux $(\Phi)$ threading through a superconducting loop, hence quantized in units of $\Phi_0$ ( $\Phi_0 = h/2e = 2.07 \times 10^{-15}$ Wb):

$$\Phi = n\Phi_0, \qquad n = 0, 1, 2, \ldots \quad, \qquad (2)$$

One can artificially incorporate a phase-shift of $\pi$ in a superconducting loop, so called "$\pi$-loop". The magnetic flux states in a $\pi$-loop are governed by the following expression:

$$\Phi = (n + 1/2)\Phi_0, \qquad n = 0, 1, 2, \ldots \quad. \qquad (3)$$

The ground state, $\Phi = \pm\Phi_0/2$, represents a pair of time-reversed degenerate states spontaneously generated in the absence of an external field, corresponding to supercurrent flowing clockwise and counter-clockwise in the loop. As a function of the loop configuration, the presence and absence of the **half-integer flux quantum effect** [1] has been employed as the basis of phase-sensitive tests of pairing symmetry.

In this article, we wish to give a historical and personal account of the advent of our tricrystal phase-sensitive symmetry experiments using scanning SQUID microscope for establishing d-wave pairing symmetry in the cuprate superconductors. On the occasion of commemorating the 100[th] anniversary of the discovery of superconductivity by Kamerlingh Onnes, we are honored to present this article as part of the Jubilee Celebration.

## 1.1 The Design of the Tricrystal Experiments

*Historical Remarks*

The tricrystal experiments were conceived in the early Spring of 1993. At the five-day 1993 APS March Meeting in Seattle, there was furious debate over the pairing symmetry in the high temperature superconductors. The root of the problem was that conventional techniques such as quasiparticle tunneling, NMR, ARPES, and penetration depth,  - - etc. can only provide information about the magnitude of the pair wavefunction, but not its phase.  At this meeting Dale Van Harlingen presented his exciting results using SQUID interference measurements in YBCO/Pb SQUIDs and junctions to support d-wave pairing symmetry (see the article by Van Harlingen elsewhere in this book). These preliminary experiments were highly controversial, so much so that the need for alternative phase-sensitive tests of pairing symmetry was clear to many attending the Meeting. During the flight back to New York, Chang Tsuei (CT) thought very hard about how to do a definitive phase-sensitive pairing symmetry experiment. By the time landed at the airport in NYC, he had formulated a basic tricrystal configuration for testing d-wave symmetry in cuprate superconductors. The crystallographic orientations of the three crystals were deliberately arranged to create a phase shift of $\pi$ in a superconducting loop, interrupted by three grain-boundary weak links around the tricrystal meeting point.  The tricrystal idea was a natural reflection of CT's experience in studying critical current density in bicrystal YBCO films[2,3] and in dc SQUIDs, made with two grain-boundary Josephson junctions, which resulted an US Patent in1988[4]. Shortly after the APS March Meeting, CT was asked to present his plan for the tricrystal experiment at a group meeting, called by Mark Ketchen, a senior manager of the Physical Sciences Department of the T.J. Watson Research Center. According to Ketchen's Lab Notebook, the meeting was named ***"S-wave d-wave superconductivity".*** Although the response to CT's presentation was quite positive and supportive, there were also strong reservations concerning whether it was feasible to make the proposed tricrystal $SrTiO_3$ (STO) substrates (needed for growing epitaxial cuprate films) with the three precisely-oriented grains, and three atomic-scale sharp grain boundaries between them. At that time, high-quality bicrystal STO substrates were commercially available for growing (001) tilt grain boundary junctions. No one had any experience or expertise of synthesizing the needed tricrystal substrates. CT ignored the skeptics and chose to work with Sinkosha Co. in Tokyo, a bicrystal STO manufacturer. The Company was willing to give it a try, with no guarantee on the quality of their final product. It turned out that the tricrystal substrates we received from Sinkosha were of truly exceptionally high-quality. The grain boundaries in the substrates were so sharp that they could not be inspected without the aid of an optical microscope equipped with polarized light.  Furthermore, all other specifications were met as well.

With this initial technical hurdle overcome, the tricrystal pairing symmetry experiment [5] was underway in earnest. As one would say: ***The rest is just history!***

*Key Considerations of the Tricrystal Design*

Phase-sensitive pairing symmetry tests are based on two macroscopic phase coherence phenomena: **flux quantization** and **Josephson pair tunneling.** Theoretical suggestions for such a probe of unconventional gap symmetry had been reported in the literature many years before. Geschkenbein, Barone and Larkin suggested half-flux quantization could result from an order parameter with nodes in polycrystalline heavy fermion superconductors in 1987 [6]. It is interesting to note that this important theoretical work was not extended to its application in cuprate high-temperature superconductors, although it was published soon after the discovery of Bednorz and Mueller in 1986. Sigrist and Rice argued that the observed paramagnetic shielding effects in granular cuprate superconductors could be due to sign changes in a d-wave pair wavefunction and proposed a SQUID interference experiment to verify such a nodal gap structure in 1992 [7]. This theoretical proposal was exactly realized with a Pb-YBCO SQUID by D. Wollman et al. in 1993 [8].

## Box 1:

### *Phase-sensitive tests of pairing symmetry*

The basic idea of phase-sensitive pairing symmetry tests of unconventional (non-simple s-wave) superconductors had existed in the literature for some time [6,7]. Due to the nature of unconventional pairing symmetry such as in the $d_{x^2-y^2}$-wave pair state, one can configure a Josephson junction (between i and j, of which at one is an unconventional superconductor) that is characterized by a negative supercurrent $I_s$:

$$I_s = -|I_c|\sin(\Delta\varphi_{ij}) = |I_c|\sin(\Delta\varphi_{ij} + \pi) \qquad (4)$$

This remarkable effect of negative pair tunneling can be as an intrinsic phase shift of $\pi$ added to the phase difference $\Delta\varphi_{ij}$ of the junction.

In a superconducting singly-connected loop containing such junctions, the phase shift is gauge-invariant and the single-valueness of the pair wavefunction $\Psi(\mathbf{k})$ demands that phase can vary only in multiples of $2\pi$ in going around a closed contour, leading to the phenomenon of flux quantization:

$$\Phi_a + I_s L + \frac{\Phi_0}{2\pi}\sum_{ij}\Delta\varphi_{ij} = n\Phi_0 \qquad (5)$$

where the self-inductance is *L*, and n is an integer. One needs to count the number of sign changes in the loop to distinguish a π-loop [half-integer flux quantization see Eq. (3)] and 0-loop [standard integer flux quantization, see Eq. (2)]. The odd-numbered sign changes cost energy in the Josephson coupling across the junctions. This excess energy is reduced by the spontaneous generation of circulating currents, which result in a Josephson vortex with a half of the flux quantum **Φ₀**,

$$U(\Phi,\Phi_a) = \frac{\Phi_0^2}{2\pi}\left\{\left(\frac{\Phi+\Phi_a}{\Phi_0}\right)^2 - \frac{L|I_c|}{\pi\Phi_0}\cos\left[\left(\frac{2\pi\Phi}{\Phi_0}\right)+\theta\right]\right\}; \quad \theta = 0, \pi \tag{6}$$

The applied flux $\Phi_a$ is zero in the ground state.

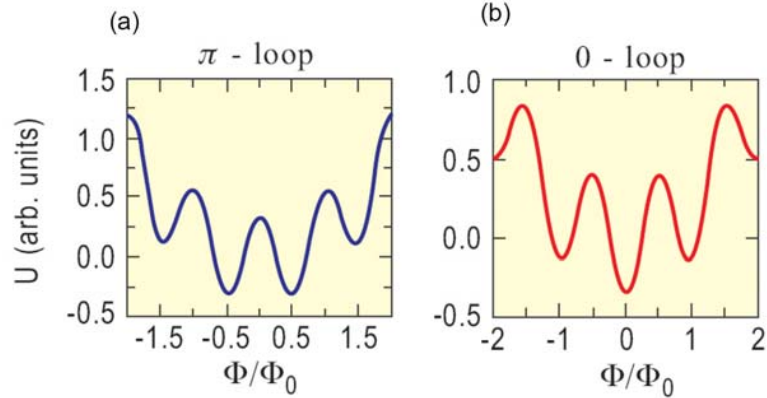

Fig.1 , Free energy of a superconducting π-loop (a) and 0-loop (b).

The doubly degenerate half flux quantum states as shown in the above figure, correspond to the supercurrent circulating in the loop clockwise and counter-clockwise respectively.

As a function of the loop geometry, the observation or non-observation of the half flux quantum effect can be used to probe the phase of the order parameter, $\Delta(\mathbf{k})$ (i.e. $\Psi(\mathbf{k})$).

The first tricrystal symmetry experiments were initiated at the IBM T.J. Watson Research Center in 1993. Key considerations that went into their design were:

*(1) Tricrystal geometry designed with Sigrist-Rice and maximum disorder formula:*

For testing for $d_{x^2-y^2}$-wave pairing symmetry, the sign of $I_S$ is determined by the Sigrist-Rice formula (i.e. in the clean limit where the junction interface is assumed perfectly smooth) [7]:

$$I_s^{ij} = \left[|I_c^{ij}|\cos 2\theta_i \cos 2\theta_j\right]\sin(\Delta\varphi_{ij}) \tag{7}$$

Where $\Delta\varphi_{ij}$ is the phase difference between grains $i$ and $j$, and $\theta_i$ and $\theta_j$ are the angle of the crystallographic axes (100) in the grains i and j with respect to the junction interface $GB_{ij}$.

To take into account of the microstructural disorder effects arising from defects such as impurities micro-facets, - - -) on tunneling normal to the junction interface, a maximum disorder (angular deviations up to a maximum of π/4 in a d-wave configuration) formula is derived for $I_S$ [5]:

$$I_s^{ij} = |I_c^{ij}|\cos 2(\theta_i + \theta_j)\sin(\Delta\varphi_{ij}). \tag{8}$$

With the combined constraints imposed by Eqs. (7) and (8), the three-grain-boundary-junction ring shown in Fig. 2a was designed to be a π-ring if the cuprate under test is a superconductor with $d_{x^2-y^2}$-wave pairing symmetry. The misorientation angles $\alpha_{12}$, $\alpha_{31}$ and the angle between the grain-boundary planes ($\beta$) were chosen to ensure that the net sign of the supercurrent in the three-junction ring is negative as required by the d-wave pair state [1, 5]. The design point of the first tricrystal experiment corresponds the solid dot in the light yellow regions of Figs. 2b and 2c.

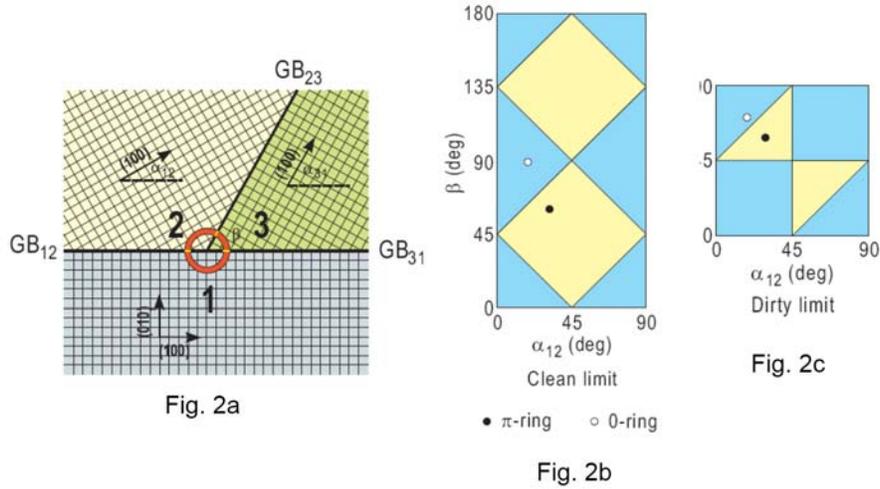

Fig. 2, (a) tricrystal configuration with design parameters defined. Design parameter regions for π- and 0-rings in the clean limit (b), and dirty limit (c).

Based on flux quantization and energy considerations, such a π-ring in the ground state should exhibit spontaneously generated magnetization of a half flux quantum, regardless of whether the grain boundaries are in the clean or dirty limit. This makes the tricrystal experiment a ***yes-or-no*** test of *d*-wave pair symmetry in terms of observing the half flux quantum or not.

*(2) Symmetry-independent mechanisms for π-phase-shift:*

Several theoretical studies have predicted that magnetic interactions (including the spin-flip scattering due to impurities [9] at the junction can induce a π-phase-shift [9, 10] in the tunnel barrier region, hence the name π-junction. In particular, such π-junctions have been demonstrated in various superconductor/ferromagnet/superconductor (SFS) thin-film structures thru the exchange-field induced spatial oscillation in the order parameter in the junction barrier [11], for a review of recent developments in SFS junctions and their potential in device applications, see [10].

A superconducting loop containing an odd number of such π-junctions can also exhibit the half flux quantum effect. It is important to differentiate between the magnetically driven π-junction and the π-loop derived from the d-wave pairing symmetry observed in the original tricrystal experiment [5]. Using the criteria of the Sigrist-Rice and maximum disorder formula, Eqs. (7) and (8), a second tricrystal experiment was designed by CT to rule out any symmetry-independent mechanism for the observed π-phase-shift effect. The design point of this tricrystal geometry is shown in Figs.2b and 2c as an open circle. Since the magnetic impurity effect would not be sensitive to such small variation in the tricrystal configuration as shown in Fig 2b and 2c. A half flux quantum would be observed in both tricrystal experiments if the spin-flip mechanism was in operation, while only in the first tricrystal experiment for the case of *d*-wave pairing symmetry.

*(3) Experimental verification of the tricrystal design parameters:*

X-ray diffraction measurements were carried to ensure a single-phase, high-quality c-axis epitaxial film growth of cuprate films e.g. YBCO on the tricrystal STO substrate with the geometry depicted in Fig.3(a). In-plane scanning x-ray diffraction and electron backscattering measurements demonstrated strong in-plane alignment in the tricrystal cuprate film with the misorientation angles at each grain boundary within 4 deg of the intended design angle. The IV characteristics of each grain-boundary junction in the tricrystal ring were shown to be those of a Josephson weak-link. The misorientation angles the tricrystal configuration in Fig.3(a) were designed to be identical to avoid any extrinsic effect due to the difference between the three junctions. The IV measurements indicated the $I_c$ values of the three boundary junctions were in agreement within 20%.

## 1.2 Direct Observation of the Half Flux Quantization

February 1994 a c-axis oriented epitaxial YBCO film was deposited on the d-wave-testing STO substrate, and micro-patterned into four rings as shown in Fig. 3(a). The three-junction ring at the center was surrounded by three 0-ring (with an even number of grain-boundary junctions) were designed to serve as the controls of the experiment. The basic design parameters of all the rings were thoroughly checked and we were finally ready to do the symmetry test. It was a great fortune that John Kirtley (JK) joined the tricrystal experiment with his scanning SQUID microscope (SSM), see Fig. 4 and Box 2).

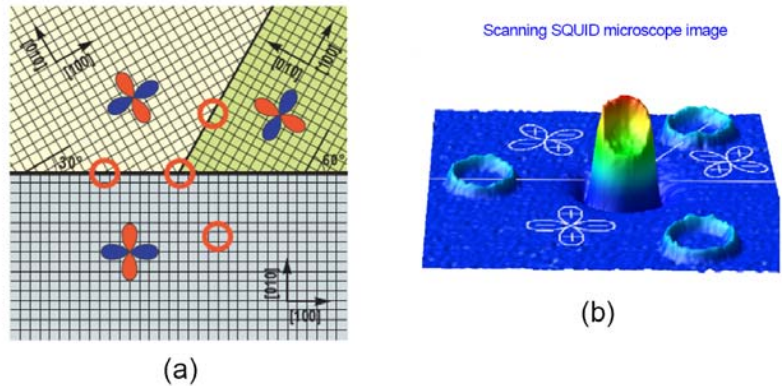

Fig 3 (a) schematic of the d-wave testing tricrystal geometry, (b) SSM image of YBCO rings at 4.2K and nominal zero field.

**Box 2:**

## *Development of a high-resolution Scanning SQUID Microscope*

The development of the scanning SQUID microscopes used for pairing symmetry tests at Yorktown Heights began in early 1992. At first this effort had nothing to do with pairing symmetry tests: Tunneling measurements in the high-$T_c$ cuprates often show a linear dependence of the dynamic conductance dI/dV on voltage. There have been a number of possible explanations for this so-called "linear conduction background", but Doug Scalapino and JK proposed that it could be due to inelastic tunneling mediated by a broad, flat distribution of excitations in the tunneling region. A review of the literature found that linear conductance backgrounds were common, including in Cr-$CrO_x$-Pb planar tunnel junctions. Al-$AlO_x$-$CrO_x$-Pb junctions had large linear conduction backgrounds, with the largest effects occurring for a few angstroms of Cr evaporated on an aluminum film before oxidation. Since a possible inelastic excitation causing the linear conduction background is spin fluctuations, it would be interesting to compare the magnetic properties of these films with the size of the linear conductance background. However, detecting magnetism in such thin films is difficult. A conventional SQUID magnetometer is not sensitive enough for this application, but a scanning SQUID system, in which the SQUID is much closer to the local dipole moments, could work. Twenty years previously IBM researchers had built a SQUID microscope[1] for imaging trapped vortices in superconducting circuitry as part of IBM's Josephson computer effort. This system no longer existed, so one was built from scratch, with Mark Ketchen designing and overseeing the fabrication of the SQUIDs, and JK building and running the microscope. JK converted an existing point contact tunneling probe[2], which used a mechanical differential thread and spring mechanism for the approach, as well as a long piezoelectric tube for the scanner, into a scanning SQUID microscope.

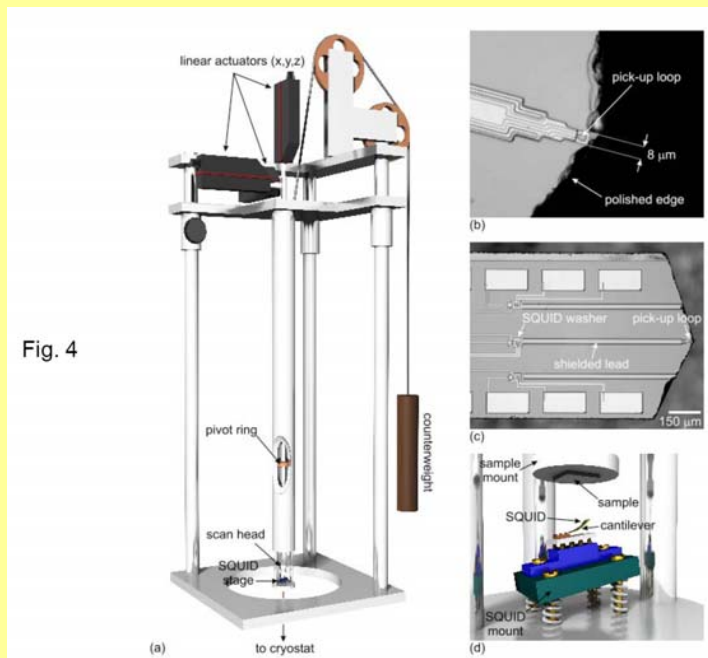

Fig. 4 (a) configuration of a scanning SQUID microscope, (b) micrograph of a scanning SQUID pick-up loop structure situated on a polished Si-wafer tip. (c) micrograph of the thin-film scanning SQUID sensor, (d) assembly of the SQUID sensor, sample, and associated mounting structures.

We first used SQUIDs connected to small superconducting pickup loops through superconducting wire bonds, then SQUIDs with pickup loops integrated into them through strip line leads, and finally SQUIDs with pickup loops integrated through coaxial sheaths. It soon became apparent that piezo tube scanning was not a good choice for a SQUID microscope, both because the largest scan range attainable was too small, but also because the large voltages associated with piezo scanning caused arcing, which destroyed the SQUIDs. JK replaced the differential thread and spring approach of the original microscope with a lever connected at one end to a 3-axis optical stage and linear actuators at room temperature, with the sample mounted on the other end. At first this was intended to be the coarse positioning mechanism, while still using piezoelectrics for the fine scanning. However, the lever mechanism worked so well that the piezotube scanner was removed. The resulting microscope[3] could scan an area about 400 microns on a side. It was never used for measuring the magnetic properties of $AlO_x$-$CrO_x$ films, in part because of the success of its application to the tricrystal experiments.

[1] F.P. Rogers, "A device for experimental observation of flux vortices trapped in superconducting thin films", Master's thesis, Massachusetts Institute of Technology, Cambridge, Massachusetts (1983).
[2] A.P. Fein, J.R. Kirtley, and R.M. Feenstra, "Scanning tunneling microscope for low-temperature, high magnetic field, spatially resolved spectroscopy", Rev. Sci. Instrum. **58**, 1806 (1987).
[3] J.R. Kirtley, M.B. Ketchen, K.G. Stawiasz, J.Z. Sun, W.J. Gallagher, S.H. Blanton, and S.J. Wind, Appl. Phys. Lett. **66**, 1138 (1995).

There are several advantages of using SSM over the SQUID interference technique for determining the amount of magnetic flux threading through a superconducting loop. First, it provides a **direct** imaging of the magnetic flux state in the ring without relying on interpreting the measured SQUID characteristics [8]. Due to the small coupling between the SSM's pickup coil and the cuprate ring, the magnetic flux ground state of the rings can be probed precisely and non-invasively. With our SSM, any spurious flux trapped on the ring can be detected to a level of $10^{-3}$ $\Phi_0$. Figure 3(b) shows a SSM image, taken at 4.2 K and nominal zero field, of four YBCO rings on a tricrystal STO substrate with the configuration depicted in Fig. 3(a). We were all very excited that flux in the 3-junction ring at the tricrystal point but not in the three control rings, as expected if YBCO were a d-wave superconductor. However, we needed to be sure that the magnitude of the flux through the 3-junction was exactly a half flux quantum, $\Phi_0/2$.

Hence, a quantitative determination of the flux state of the rings became the crucial part of the tricrystal experiment. We eventually used a number of flux calibration techniques. The most straightforward one was direct calculation of the flux through the pickup loop due to currents in the rings. The flux $\Phi_p$ through the pickup loop can be written as $\Phi_p=\Phi_r M/L$, where $\Phi_r$ is the flux in the ring, M is the mutual inductance between the pickup loop and the ring, and $L$ is the self-inductance of the ring. Calculating $M$ is straightforward, but estimating $L$ is more difficult, since field penetration into the superconducting body of the ring must be carefully accounted for. Nevertheless, Mark Ketchen had previous experience with this type of problem, and his initial estimates for $L$ were within 10% of the real value. CT vividly recalled that Mark called CT at home around one o'clock in the morning to discuss with him the L value he obtained. They concluded that it was probably a half flux quantum threading thru the 3-junction ring. John C.C. Chi and Jonathan Sun also provided their calculated L values using different approximations. All these calculations agreed within 5%., and very convincingly led us to the conclusion that it was a $\Phi_0/2$=h/4e flux trapped in the 3-junction ring!

We nicknamed the second calibration technique the "oil drop" method: we cooled the rings repeatedly in various fields and measured the difference between the SQUID signal when it was centered above a ring vs. that away from a ring. For all of the rings these differences were integer multiples of a value corresponding to a flux of $\Phi_0$, but the differences were offset by half of this value for the 3-junction ring relative to that for the control rings – exactly what one would expect if the 3-junction ring was experiencing the half-flux quantum effect while the control rings had conventional flux quantization. A third calibration technique was to apply an external field to the sample, while monitoring the flux through the SQUID with the pickup loop centered above one of the rings. Over a certain range of fields the SQUID flux jumped in discrete steps as the applied field was increased, each step corresponding to an increase in flux of $\Phi_0$ through the ring. A fourth calibration technique was to run cross-sections through the center of one of the rings as a function of externally applied field **B**$_{ext}$. The SQUID signal was the same inside the ring as outside the ring when the externally applied flux $\Phi_{ext}$=**B**$_{ext}A$, where $A$ is the effective area of the ring. The SQUID signal difference vs. **B**$_{ext}$ characteristics crossed zero at **B**$_{ext}$ values corresponding to integer multiples of $\Phi_0$, while the 3-junction ring had zero crossings offset by $\Phi_0/2$, within experimental errors of a few percent. We called this last method "titration", because the mechanism for detecting the zero crossing was not important, just as in a chemical titration. All four methods agreed, making it clear that the 3-junction rings in the first tricrystal geometry were exhibiting the half-flux quantum effect.

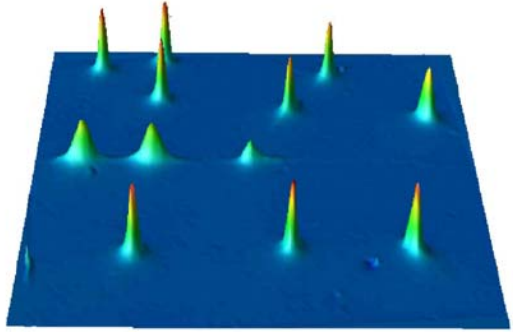

Fig. 5 SSM image of an unpatterned tricrystal c-axis epitaxial YBCO film

It was both tedious and quite challenging to lithographically pattern rings from the expitaxial cuprate films with the 3-junction ring centered on the tricrystal point. The patterned ring configuration was essential for a straight forward flux determination, without the need of modeling, in the initial tricrystal experiments. Based on the discussion earlier in this paper, a Josephson vortex with $\Phi_0/2$ of total flux is spontaneously generated at the tricrystal point for a *d*-wave superconductor epitaxially grown on a tricrystal substrate with the appropriate geometry as expected from the design. If such a sample is cooled in a zero field, only the half-flux quantum Josephson vortex should present. Upon cooling in a finite field, Abrikosov vortices in the bulk, and integer Josephson vortices along the grain boundaries should be also observed. Careful integration of the flux, taking into account the SSM-and-sample geometrical configuration, showed that the vortex at the tricrystal point had half of the flux of the other single vortices (see Fig. 5) [12]. The technique of using unpatterned tricrystal epitaxial cuprate films is especially well-adapted for testing pairing symmetry in various cuprate superconductors other than YBCO.

## 1.3 Elucidation of the Nature of Half Flux Quantum Effect

After the first tricrystal experiment, a series of phase-sensitive symmetry experiments were carried out to elucidate the nature of the observed half flux quantum effect:

*(1) Symmetry-independent mechanism for π-phase-shift:*

Right after we published the results of the first tricrystal experiment, there were several suggestions to us that the half flux quantum effect we reported might be due to the impurity and defects at the grain boundary of our tunnel junctions through the spin flip scattering or electronic correlation effect. In response these criticisms, CT designed a new tricrystal experiment to demonstrate that by a small variation in the tricrystal configuration (the design point in the 0-loop regime as shown in Fig. 2b) we could turn **on** and **off** the half flux quantum effect in accordance with the d-wave pairing symmetry assumption [Eqs.(7) and (8)]. As shown in Fig. 6(b), the SSM image of the new three-junction ring and the controls exhibited no flux in their ground state [13]. This would not happen if any symmetry-independent mechanism was in operation.

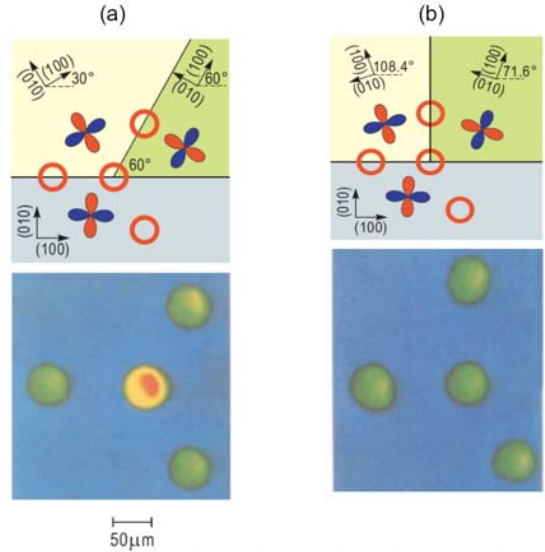

Fig. 6, tricrystal geometrical configurations and SSM images of rings:(a) designed to be a π- ring at the tricrystal point for a d-wave superconductor, (b) 0-ring designed to rule out any symmetry-independent mechanism for the half flux quantum effect.

*(2) Bilayer effect in cuprate superconductors:*

In the early stage of determining d-wave pairing symmetry in cuprates, all phase-sensitive experiments were done with YBCO whose crystal structure is characterized by two CuO2 planes per unit cell. It was suggested that the two Cu-O layers can carry two s-wave order parameters but with opposite signs and mimic the observed $d_{x^2-y^2}$ –wave gap function [14, 15]. This issue was resolved by a d-wave pairing symmetry test with single-layer tetragonal $Tl_2Ba_2CuO_{6+\delta}$ (Tl2201) superconductors [16]. The observation of spontaneously generated half flux quantum at the three-junction ring centered at the tricrystal meeting point is a convincing evidence to rule out the bilayer effect. The half flux quantum effect was also observed in the tricrystal blanket film of Tl2201.

*(3) Tetracrystal pairing symmetry experiment:*

All the phase-sensitive pairing symmetry experiments with YBCO can not distinguish between pure d-wave pairing and a *d*-wave dominant *d+s* mixed pair state. The order parameter should transform in accordance with the symmetry operations of the relevant crystal point group. In an orthorhombic superconductor such as YBCO, *s*-wave and *d*- wave spin-singlet pairings correspond to the same irreducible representation $A_{1g}$ of the point group $C_{2v}$. A *d+s* mixed pair state is expected. For a tetragonal single-layer Tl2201 cuprate superconductor with point group $C_{4v}$, *s*-wave and *d*-wave pairings correspond to the irreducible representation $A_{1g}$ and $B_{1g}$ respectively. An admixture of *s*-wave and *d*-wave pairings is not allowed. To test these notions a tetracrystal experiment with a π/4-rotated wedge geometry was designed and implemented. The interpretation of this experiment depends only on symmetry arguments and is therefore independent of any model for the Josephson pair tunneling current (i.e. Eqs. 7 and 8). The observation of the half-flux quantum effect in the tetracrystal experiment represents a strong and model-independent evidence for pure and $d_{x^2-y^2}$–wave pairing state in cuprates with tetragonal crystal structure [17]. This work also served as the basis of the later development for all d-wave dc SQUIDs [18] and related superconducting device applications [19-21].

## 1.4 Universality of the $d_{x^2-y^2}$ pair state

In the past, numerous theoretical studies suggested that the stability of the d-wave pair state in various cuprates could be significantly affected by the details in the band structure and pairing potential. Especially the competing nature of the s-wave and d-wave pairing channels was emphasized. To clarify these issues, several phase-sensitive tricrystal experiments were carried out:

*(1) Doping effect on pairing symmetry in hole-doped cuprates:*

High-temperature superconductivity in cuprates is achieved by doping the Mott-insulators with charge carriers, electrons and holes. A definitive determination of doping effect on pairing symmetry in cuprate superconductors is important in the study of high-$T_c$ mechanism and quantum critical phenomena. We have done systematically a series of tricrystal experiments to study the doping dependence of gap symmetry in a variety of hole-doped cuprate superconductors. By an appropriate heat treatment of the $Bi_2Sr_2CaCu_2O_{6+\delta}$ system, tricrystal experiments were carried out at the optimum doping and to scan the pairing symmetry over a wide range of doping covering both the over- and under-doped regimes. Also included in the study were $La_{2-x}Sr_xCuO_4$, a historically and fundamentally important high-$T_c$ superconducting system, Tl2201system, optimally doped YBCO and Ca-doped YBCO system. Our results indicate that the *d*-wave pairing persists through the entire phase diagram [22].

*(2) Electron-doped cuprates:*

It turned out to be quite difficult to do a tricrystal experiment with the electron-doped cuprate superconductors $Nd_{1.85}Ce_{0.15}CuO_{4-y}$ and $Pr_{1.85}Ce_{0.15}CuO_{4-y}$. This is mainly because that the critical current density $J_c$ at the grain boundary with 30° misorientation angle (as shown in Fig. 2a) is five orders of magnitude smaller than that of YBCO. As a result, the spatial extension of half flux quantum signal, as measured by $2\lambda_J$ ~100μm (the Josephson penetration depth $\lambda_J \propto J_c^{-1/2}$), is much more broader than in the YBCO case.

Despite the weak SSM signal at the tricrystal point, we managed to demonstrate and use the presence and absence of half flux quantum effect in a series of phase-sensitive tricrystal experiments [23] designed to be π- and 0-loops at the tricrystal point for a *d*-wave superconductor (see Fig.7). These results represent strong evidence that the electron-doped cuprates are *d*-wave superconductors as their hole-doped counterparts. In addition, there was convincing evidence that the time reversal symmetry is also conserved in both e-doped superconductors.. It is remarkable that, although their normal-state properties are quite different from those of their hole-doped counterparts, the d-wave pairing dominates in all cuprates.

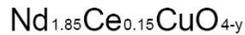

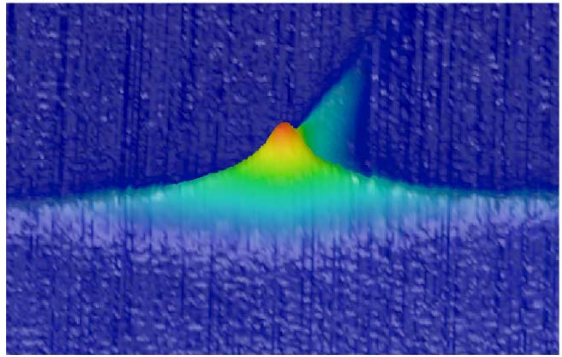

Fig. 7 Tricrystal pairing symmetry tests of electron doped cuprates

*(3) Temperature dependence of the half flux quantum effect:*

Although the tricrystal experiments were widely accepted to provide conclusive evidence for *d*-wave superconductivity in the cuprates at liquid helium temperatures, there was speculation that the pairing symmetry could change with temperature, with the inclusion of a real or imaginary sub-dominant component at either low or high temperatures. We therefore built a scanning SQUID microscope with the capability of varying the sample temperature over a large range. Temperatures down to 0.5K were achieved by condensing $^3$He in the sample space and pumping on it; temperatures up to 100K were attained by evacuating the sample space, providing a high thermal conductivity link between the SQUID sensor and the $^4$He bath, and heating the sample. Measurements of a YBCO tricrystal sample using this microscope showed that the half-flux quantum vortex had, to within experimental error of a few percent, $\Phi_0/2$ total flux over a temperature range from 0.5K to within a few degrees of $T_c$=90K [24]. This indicated that there was no change in the dominant *d*-wave symmetry, and little if any imaginary component in the order parameter, over the entire temperature range.

In brief summary, the results of a systematic study using tricrystal experiments have demonstrated that the d-wave pair state in cuprates is robust against large varations in temperature up to T_c and doping covering both the under- and over-doped regimes. This finding underscores the important role of strong onsite Coulomb interaction, a universal characteristic of all cuprates, in favoring the d-wave over the s-wave pairing channel. When the strong correlation condition is relaxed, as presumably happened in the recently discovered Fe-based pnictides and tellurides, the s-wave pairing can prevail. In fact, due to the multi-orbital nature of these Fe-based high-temperature superconductors, a nodeless s-wave order parameter with sign reversal is also possible [25, 26].

## 1.5 Large-scale arrays of the half flux vortices

The tricrystal and tetracrystal experiments described so far were all characterized by one single half flux quantum in a π-ring per sample. To gain insight in the gap structure of YBCO, for example, and for paving the road to future device applications in quantum computing and so on, one needs the capability of fabricating large-scale integration of arrays of π-loops on one single wafer. Two of such efforts are presented below:

*(1) Two-dimensional arrays of π-loops*

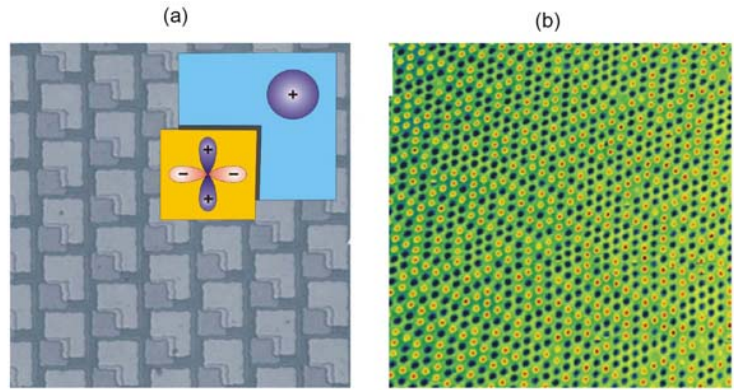

Fig. 8  Two-dimensional triangular π–arrays, (a) SEM micrograph, (b) SSM micrograph of the flux state in a triangular lattice.

A large-scale integration of superconducting π-loops on a single wafer was made possible due to the development of a process, by the H. Hilgenkamp and D. Blank groups of the University of Twente in the Netherlands, for making high quality ramp-edge Josephson junctions between YBCO and Nb[27]. This process used an in-situ etch and re-growth of the YBCO after photolithographic definition of the ramp edge that allowed for junction interfaces for making Josephson junctions with high critical current density and high reproducibility. In collaboration with the Twente group, fabrication of a triangular lattice 25000 π-loops on a single chip, of size 5mm x 10mm, was demonstrated. We were able to use the SSM to image and to study the ordering and manipulation of the half-flux quantum states in such large arrays [28] (see Figs.8a and 8b). This work opens the door to fundamental studies, including phase transitions in geometrically frustrated non-frustrated model systems (2D-Ising models), and possible superconducting device applications [19] .

*(2) Angle-resolved determination of gap anisotropy in YBCO*

After doing all the tricrystal and tetragrystal experiments described above, there was one important unresolved issue in pairing symmetry. Group theory says that a d+s mixed pair state in YBCO is unavoidable due to its orthorhombic crystal symmetry **C$_{2v}$**. However, the size of the gap anisotropy was a highly controversial topic for many years. Early attempts to measure the in-plane momentum dependence of the order parameter in YBCO and other cuprates using tunneling measurements had limited success. Even the phase-sensitive experiments, for example the interference using a Pb-YBCO corner or single Josephson junction, can only conclude qualitatively that d > s, i.e. the s-wave component in the s+d pairing admixture is less than 50%. This is due to the fact that these experiments were based only on a sign change in order parameter between the orthogonal a and b faces of a YBCO single crystal, and therefore do not provide any quantitative information about the d/s ratio.

A golden opportunity for solving this problem arose when Hans Hilgenkamp visited IBM in Yorktown after he attended the APS March Meeting 2004 in Montreal. It was at a meeting of HH, JK and CT at the Cafeteria of the Watson Research Center, when CT presented his proposal for making a series of two-junction Nb/YBCO loops to map out quantitatively the in-plane gap anisotropy in YBCO. The proposed experiment demanded the fabrication of many high-quality 2-junction rings with the second ramp-edge junction angle varying, from ring to ring, in increments of a few degrees and less. Furthermore, the critical currents of all the ramp-edge junctions should be large enough for a viable study of the magnetic flux state in about seventy 2-junction rings.

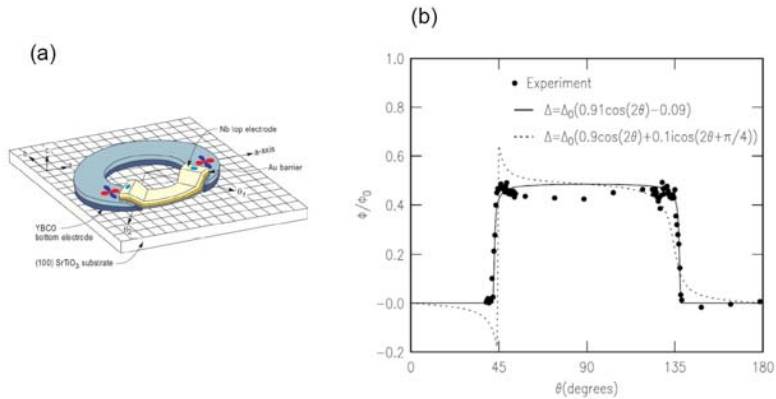

Fig. 9 Angle resolved phase-sensitive determination of gap in YBCO. (a) schematic of the 2-junction ring, (b) angular dependence of integrated flux thru the rings.

Even with the prior experience of fabricating large-scale π-arrays [28], this experiment represented a tremendous challenge to HH and his colleagues. Several months later, we were pleased to receive a set of beautiful arrays of the 2-junction YBCO/Nb rings, see Fig. 9(a). One of the ramp-edge junction angles relative to the *a*-axis direction was held fixed at -22.5°. The second junction angle was varied in 5° intervals between -107.5° and 242.5°. There were a total of 72 two-junction rings on the same chip!! HH and his team took great care optimizing the ring design to ensure that the rings would have the required uniformity (among all the rings on the same chip) and large $I_cL$ products of each individual ring for observing well-defined flux quanta.

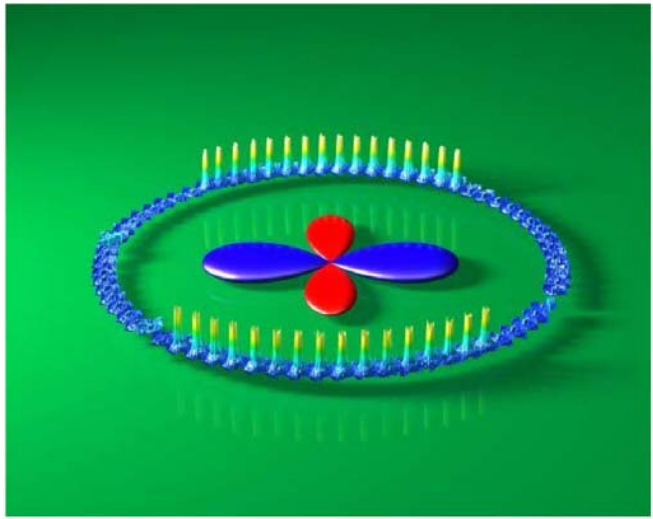

Fig. 10  A polar plot of SQUID microscope images of 72 rings with different angular geometries for determining the in-plane gap anisotropy in YBCO.

JK analyzed his SSM results on the first sample and we concluded that the data indicated the nodal transition occurred at an angle deviated from 45°, the angular position for the pure $d_{x^2-y^2}$ gap. While we were enjoying the promising data from the first sample, it was clear that we needed a second sample with the second junction angle to vary in 0.5° intervals in the nodal regions around 45° and 135°. However, our project suffered a "phase-shift" because a team member, Ariando, at Twente graduated with his PhD. At that point, a new graduate student, C.J.M. Verwijs joined the team. HH and CT worked together to get him to phase-in quickly and he managed to produce "super chips" that allowed us to fill in the data points in the nodal transition region with angular resolution in the second angle better than 0.5° ! See Figs. 9 (b) and Fig. 10. The combined results of the two samples were fit by JK with a functional form $\Delta(\theta)=\Delta_0(0.91\cos(2\theta)-0.09)$ suggesting that the gap along the b-axis of YBCO is at least 20% larger than that along the a-axis direction [29]. The results also showed that any imaginary component to the order parameter, if present, must be quite small.

# Concluding Remarks

In summary, building on the results obtained with conventional gap-amplitude measurements (quasiparticle tunneling spectroscopy, NMR, ARPES, and penetration-depth studies), the development of phase-sensitive techniques has finally settled the decade-long d-wave versus s-wave debate in favor of an order parameter with $d_{x^2-y^2}$ symmetry in both hole- and electron-doped cuprate superconductors. In this article, we recall our experience of doing the scanning SQUID microscope phase sensitive experiments, using the half-flux-quantum effect as the definitive signature for establishing d-wave pairing symmetry. Through a series of tricrystal experiments, we have demonstrated that the d-wave pair state is robust against a wide range temperature and doping variations and time-reversal symmetry breaking.

The journey to d-wave has been very exciting and rewarding for us. Along the way, we were supported and encouraged by many colleagues. Without their contributions, the work we described here would have been impossible. Here, we wish to thank a few of them: Ariando, C. C. Chi, W. J. Gallagher, H. Hilgenkamp, R. P. Huebener, M. B. Ketchen, D. H. Lee, Z. Z. Li, J. Mannhart, K. A. Moler, D. M. Newns, H. Raffy, Z. F. Ren, J.Z. Sun, G. Trafas, C. J. M. Verwijs, M. B. Walker, Z. H. Wang and S. K. Yip.

We would like to end our article with Maurice Rice's comment in AIP Research Highlights of the year 1994: *" - - - -. These experiments are an important milestone on the way to a complete microscopic theory of this spectacular phenomenon. Only when a reliable formula for $T_c$ with predictive power has been obtained can one say the journey is ended"*.

# References

[1] C.C. Tsuei and J.R. Kirtley, Rev. Mod. Phys. **72**, 969 (2000).

[2] P. Chaudhari, J. Mannhart, D. Dimos, C.C. Tsuei, J. Chi, M.M. Oprysko, and M. Scheuermann, Phys. Rev. Lett. **60** 1653 (1988).

[3] J. Manhart, P. Chaudhari, D. Dimos, C.C. Tsuei, and T.R. McGuire, Phys. Rev. Lett. **61**, 2476 (1988).

[4] P. Chaudhari, C.C. Chi, J. Mannhart, and C.C. Tsuei "Grain Boundary Junction Devices Using High Tc Superconductors", US Patent 5162298.

[5] C.C. Tsuei, J.R. Kirtley, C.C. Chi, Lock See Yu-Jahnes, A. Gupta, T. Shaw, J.Z. Sun, and M.B. Ketchen, Phys. Rev. Lett. **73,** 593 (1994).

[6] V.B. Geshkenbein, A.I. Larkin, and A. Barone, Phys. Rev. B **36**, 235 (1987).

[7] M. Sigrist and T. M. Rice, J. Phys. Soc. Japan, **61**, 4283 (1992).

[8] D.A. Wollman, D.J. Van Harlingen, W.C. Lee, D. M. Ginsberg, and A. J. Leggett, Phys. Rev. Lett. **71**, 2134 (1993).

[9] L.N. Bulaevskii, V.V. Kuzii, A.A. Sobyanin, JETP Lett. **25**, 290 (1977).

[10] Hans Hilgenkamp, Supercond. Sci. Technol. **21**, 034011 (2008).

[11] V.V. Ryazanov, V.A. Oboznov, A.Yu. Rusanov, A.V. Veretennikov, A.A. Golubov, and J.Aarts, Phys. Rev. Lett. **86**, 2427 (2001).

[12] J.R. Kirtley, C.C. Tsuei, M. Rupp, J.Z. Sun, Lock See Yu-Jahnes, A. Gupta, and M.B. Ketchen, Phys. Rev. Lett. **76**, 1336 (1996).

[13] J.R. Kirtley, C.C. Tsuei, J.Z. Sun, C.C. Chi, Lock-see YuJahnes, A. Gupta, M. Rupp, and M.B. Ketchen, Nature **373**, 225 (1995).

[14] D.Z. Liu, K. Levine, J. Malay, Phys. Rev. B **51**, 8680 (1995).

[15] A.I. Liechienstein, I.I. Mazin, O.K. Andersen, Phys. Rev. Lett. **74**, 2303 (1995).

[16] C.C. Tsuei, J.R. Kirtley, M. Rupp, J.Z. Sun, A. Gupta, M.B. Ketchen, C.A. Wang, Z. F. Ren, J.H. Wang, M. Bhushan, Science **271,** 329 (1996).

[17] C.C. Tsuei, J.R. Kirtley, Z.F. Ren, J.H. Wang, H. Raffy, and Z.Z. Li, Nature **387**, 481 (1997).

[18] R.R. Schulz, B. Chesca, B. Goetz, C.W. Schneider, A. Schmehl, H. Bielefeldt, H. Hilgenkamp, J. Mannhart, and C.C. Tsuei, Appl. Phys. Lett. **76**, 912 (2000).

[19] T. Ortlepp, Ariando, O. Mielke, C.J.M. Verwijs, K.F.K. Foo, H. Rogalla, F.H. Uhlmann, H. Hilgenkamp, Science **312**, 1495 (1006).

[20] L.B. Ioffe, V.B. Geshkenbein, M.V. Felgel'man, A.L. Fauchere, and G.Blatter,

Nature **398,** 679 (1999).

[21] Thilo Bauch, Tobias Lindström, Francesco Tafuri, Giacomo Rotoli, Per Delsing, Tord Claeson, and Floriana Lombardi, Science **311**, 57 (2006).

[22] C.C. Tsuei, J.R. Kirtley, G. Hammerl, J. Mannhart, H. Raffy, Z.Z. Li, Phys. Rev. Lett. **93**, 187004 (2004).

[23] C.C. Tsuei and J.R. Kirtley, Phys. Rev. Lett. **85**, 182 (2000).

[24] J.R. Kirtley, C.C. Tsuei, and K.A. Moler, Science **285**, 1373 (1999).

[25] I.I. Mazin, D.J. Singh, M.D. Johannes, and M.H. Du, Phys. Rev. Lett. **101,** 057003 (2008).

[26] C.-T. Chen, C.C. Tsuei, M.B. Ketchen, Z.A. Ren, and Z.X. Zhao, Nature Physics **6**, 260 (2010).

[27] H.-J.H. Smilde, H. Hilgenkamp, G. Gijnders, H. Rogalla, and D.H.A. Blank, Appl. Phys. Lett. **80**, 4579 (2002).

[28] Hans Hilgenkamp, Ariando, Henk-Jan H. Smilde, Dave H. A. Blank, Guus Rujnders, Horst Rogalla, John R. Kirtley, and Chang C. Tsuei. Nature **422,** 50 (2003).

[29] J.R. Kirtley, C.C. Tsuei, Ariando, C.J.M. Verwijs, S. Harkema, H. Hilgenkamp Nature Physics **2,** 190 (2006).